\documentclass[journal]{IEEEtran}
\ifCLASSINFOpdf
\else
\fi
\let\chapter\section
\usepackage{amsmath}
\usepackage{graphicx}
\usepackage{subfigure}
\usepackage{multirow}
\usepackage{amssymb}
\usepackage{algorithmic}
\usepackage{caption}
\usepackage{color}
\usepackage[ruled,vlined]{algorithm2e}
\usepackage{bm}
\usepackage{enumerate}
\usepackage{mathrsfs}
\usepackage{leftidx}
\usepackage{authblk}
\usepackage{booktabs}
\usepackage{threeparttable}
\usepackage{array}
\usepackage{amsfonts,amssymb}
\usepackage{textcomp}
\usepackage{hyperref}
\usepackage{cite}
\usepackage{tikz}
\usetikzlibrary{positioning}
\newcommand{\tabincell}[2]{\begin{tabular}{@{}#1@{}}#2\end{tabular}}

\begin{document}
\captionsetup[table]{name={TABLE},labelsep=none}
\captionsetup[figure]{name={Fig.},labelsep=period}

\title{Novel Maximum Likelihood Estimation of Clock Skew in One-Way Broadcast Time Synchronization}

\author{
        Fanrong~Shi,~\IEEEmembership{Student Member,~IEEE},
        Huailiang~Li,~\IEEEmembership{Member,~IEEE},
        Simon~X.~Yang,~\IEEEmembership{Senior Member,~IEEE},
        Xianguo~Tuo
        and~Maosong~Lin

\thanks{Manuscript received May 20, 2019; revised September 6, 2019; accepted October 27, 2019. This work was supported in part by National Natural Science Foundation of China Programs (61601383, 41774118), Sichuan Science and Technology Program (No.2018GZ0095), and Longshan academic talent research supporting program of Southwest University of Science and Technology (17LZX650,18LZX633). (Corresponding authors: Huailiang Li and Simon X. Yang)}

\thanks{Fanrong Shi and Maosong Lin are with Robot Technology Used for Special Environment Key Laboratory of Sichuan Province, School of Information Engineering, Southwest University of Science and Technology, Mianyang 621010, China (e-mail: sfr\_swust@swust.edu.cn, lms@swust.edu.cn).}
\thanks{Huailiang Li is with the College of Geophysics, Chengdu University of Technology, Chengdu 610059, China. (email: lihl@cdut.edu.cn).}
\thanks{Simon X. Yang is with Advanced Robotics and Intelligent Systems (ARIS) Laboratory, School of Engineering, University of Guelph, Guelph, Ontario, N1G 2W1, Canada (email: syang@uoguelph.ca).}
\thanks{Xianguo Tuo is with Sichuan University of Science and Engineering, Zigong 643000, China (e-mail:tuoxg@cdut.edu.cn).}

}

\markboth{}
{Shell \MakeLowercase{\textit{et al.}}: Bare Demo of IEEEtran.cls for IEEE Journals}

\maketitle

\begin{abstract}

Clock skew compensation is essential for accurate time synchronization in wireless networks. However, contemporary clock skew estimation is based on inaccurate transmission time measurement, which makes credible estimation challenging. Based on one-way broadcast synchronization, this study presents a novel maximum likelihood estimation (MLE) with an innovative implementation to minimize the clock skew estimation error caused by delay. A multiple one-way broadcast model is developed for observation and set collection, while the distribution of actual delay is discussed. A clock skew MLE based on Gaussian delay is then proposed and its Cramer-Rao lower bound provided. The developed MLE is independent of the clock offset estimation, and requires only two synchronization periods to generate an accurate estimation. Extensive experimental results indicate that the performance of the proposed MLE is better than the methods popularly used in existing time synchronization protocols.

\end{abstract}


\IEEEpeerreviewmaketitle

\section{Introduction}
\IEEEPARstart{S}{ynchronization} is a classical problem, and many distributed systems rely on reliable synchronization, e.g., synchronization control \cite{1,2,3,5}, synchronization measurement \cite{8,9}, industrial wireless networks \cite{6,7} and wireless sensor networks (WSN) \cite{11,12}.

In recent years, wireless networks have been applied extensively in the military, industrial, commercial, and health fields. Time synchronization is an essentially important issue for wireless network applications with large-scale network nodes and variable network shapes. For instance, time synchronization requirements in the industrial automation standards ISA100.11a \cite{ISA-100.11a} and WirelessHART \cite{WirelessHART} are about $\pm100$ {\textmu s} and $\pm96$ {\textmu s}, respectively; and in structural health monitoring is 120 {\textmu s}\cite{SHM_S}. Large-scale WSN are resource-constrained and suffer from unreliable connections, which makes time synchronization a major challenge.

A real-time clock skew compensation is used in the time synchronization protocol to counteract the instantaneous clock drift among clock sources and is an indispensable technology for accurate and efficient time synchronization. Time synchronization protocols in \cite{17,18,19,20} only correct the clock offset, and the network must be re-synchronized after a very short interval to maintain high synchronization precision \cite{21}. By compensating for the clock skew and clock offset concurrently, the accuracy and efficiency of the time synchronization in\cite{22,23,24,25,26,27,28,29,30,31,32} are significantly improved. Hence, an accurate clock skew estimation is as important as the clock offset estimation for stable, and high precision time synchronization in large-scale networks.

Linear regression (LR) is a popular clock skew estimation for existing synchronization protocols, e.g. reference broadcast synchronization (RBS) \cite{22}, flooding time synchronization protocol (FTSP) \cite{23}, PulseSync \cite{24}, flooding with clock speed agreement (FCSA) \cite{25}, coefficient exchange synchronization protocol (CESP) \cite{33}. According to the implementation of LR, clock offset estimates should be obtained before regression calculations. Therefore, the clock skew estimation may not be accurate until the clock offset estimation has converged. Consequently, the synchronization convergence time depends on the table size and synchronization interval. In addition, once there is a sudden disturbance that occurs randomly in time (e.g., the sample with uncertain delay introduced in this study), it may take many synchronization periods to recover the synchronization accuracy and for that samples to be moved out from the regression table. Such samples are inevitably due to the uncertain delay caused by software blocking, interrupt handling, and interrupt nesting.

Maximum likelihood estimation (MLE) is employed to minimize the estimation error due to random variable delay. Most proposed MLEs are based on the two-way message exchange model, and the joint skew-offset estimator in which the delay is commonly modeled as Gaussian or as exponentially distributed. Jeske \cite{34} was first to provide an MLE of clock offset in a two-way message exchange model. Ahmad et al. \cite{26} presented the MLE of clock skew based on convex optimization. Subsequently, Noh et al. \cite{27} and Chaudhari et al. \cite{35} removed the nuisance parameters and proposed a joint MLE of clock skew and offset estimation. Li and Jeske \cite{36} obtained the MLE of clock offset and clock skew using re-parameterization to simplify the form of the likelihood function. However, if the time synchronization algorithm based on the two-way message exchange cannot avoid the relying on network hierarchy and topology management, it may fail to achieve timely and efficient time synchronization over large diameter and dynamic networks because of its pairwise synchronization mechanism.

Clock skew estimation based on the one-way broadcast model is more robust and more easily implemented but still faces major challenges. In time synchronization algorithms based on the one-way broadcast model, e.g., FTSP, FCSA, PulseSync, average TimeSync (ATS) \cite{28}, gradient time synchronization protocol (GTSP) \cite{29}, consensus
clock synchronization (CCS) \cite{30}, maximum time synchronization (MTS) \cite{31}, the LR or a direct estimator are used to estimate the relative clock skew based on the timestamps created at sender and receiver.
The Kalman filter (KF) is used to minimize the clock skew estimation \cite{KF_1,KF_3,KF_2}. However, due to poor computing ability, energy-constraint, packets loss, and clock shift in the wireless networks (e.g., WSN), the KF measurement model may be unreasonable and needs to be redesigned.
In addition, the transmission delay is the main estimation error for the clock offset estimate and clock skew estimate. Most delays can be eliminated by using start-of-frame delimiter (SFD) interrupt and medium access control (MAC) layer timestamp techniques, but the variable delay is inevitable. Furthermore, the uncertain delay, which may be caused by multiple tasks, multilevel interrupt priority level, and interrupt nesting, may cause synchronization to fail. Thus, a robust clock skew estimation in one-way broadcast should be investigated in greater depth.

In this study, a new MLE and its innovative implementation for accurate and robust clock skew estimation are developed. To minimize the estimation error resulting from the variable delay, a clock skew MLE based on the Gaussian delay model is proposed, and the Cramer-Rao lower bound (CRLB) is proposed. The implementation of the developed MLE is based on the multiple one-way broadcast model, on which, a first-in, first-out (FIFO) sliding buffer memory is designed to provide the best MLE. In addition, a $3\sigma$ detector is employed to counteract the possible sudden disturbance of delay (e.g., uncertain delay caused by interrupt nesting and queueing delay).The clock skew estimated error of MLE is about 3-4 times smaller than LR and 12-13 times smaller than direct estimation.

The proposed MLE is:

(1) A fully distributed clock skew estimation that is independent of clock offset estimation and does not require topology management. Hence it is very suitable for distributed time synchronization algorithms.

(2) A rapid convergent and accurate clock skew estimation method in which only the time information of the current and last synchronization periods is required, and it may adapt quickly to the changes in clock drift (e.g. due to variances in voltage and temperature) or networks topology changes (e.g. due to the failure node and mobile node).

(3) Easy to implement and energy efficient. Hence, it is a superior clock skew estimator for the energy-constrained and resource-constrained WSN. In addition, the proposed implementation can easily be used to develop an accurate time synchronization protocol.

(4) Able to provide a reliable clock skew estimate at different clock resolutions. It shows better performances than comparison methods when the clock granularity of the time synchronization algorithm grows rough.

We present our system model in Section II, along with the multiple one-way broadcast model. In Section III, we provide the actual experimental delay and present our MLE of clock skew. The implementation of the MLE is developed in Section IV, and the experimental results are analyzed in Section V. The conclusions are presented in Section VI.

\section{System Model and Delay Analysis}
The WSN with $\mathcal{N}$ nodes is modeled as the graph $G=(\mathcal{V},\mathcal{E})$, where $\mathcal{V}=\{1,2,\ldots,\mathcal{N}\}$ represents the nodes of the WSN and $\mathcal{E}$ defines the available communication links. The set of neighbors of node $v_i$ is $\mathcal{N}_i=\{j|(i,j)\in \mathcal{E},i\neq j\}$, where nodes $v_i$ and $v_j$ belong to $\mathcal{V}$, and $j\in \mathcal{N}_i$. Defining the hardware clock $H(t)$ and logical clock $L(t)$ as
\begin{equation}\label{equ:1}
  H(t)=\int_{t_0}^t h(\tau)d\tau,
\end{equation}
\begin{equation}\label{equ:2}
  L(t)=\varphi(t)\times {\int_{t_0}^t h(\tau)d\tau}+\theta(t_0)
\end{equation}
where $\theta(t_0)$ is the initial clock offset; and $h(\tau)$ is the clock rate of the hardware clock $H(t)$ and depends on the actual frequency of the clock source (crystal oscillator). In addition, $h(\tau)$ changes fast due to electronic noise and environmental changes, e.g., temperature \cite{38} and voltage \cite{39}. The actual clock frequency of the node is not the same as its nominal frequency. Thus, the actual $h(\tau)$ is completely different from the ideal clock rate that depends on the nominal frequency. In an actual WSN application, $h(\tau)$ cannot be measured and configured. Therefore, estimating the actual value of $h(\tau)$ is challenging and very important for the time synchronization algorithm. Parameter $\varphi(t)$ is a clock skew multiplier in software that speeds up or slows down $L(t)$, and is usually initialized to 1.

\subsection{Relative Clock Skew}
With respect to arbitrary nodes $v_i$ and $v_j$, $H_i(t)$ and $H_j(t)$, respectively, are the time notions of the nodes. Parameter $\varphi_i^j$ is used to describe the relative clock frequency rate (clock skew) of $v_i$ and $v_j$, i.e., $\varphi_i^j=h_j/h_i-1$, and it can be calculated by
\begin{equation}\label{equ:3}
  \varphi_i^j(t)=\theta _\Delta(\tau) /\tau,    \tau>0
\end{equation}
\begin{equation}\label{equ:4}
  \theta_\Delta(\tau)=(H_j(t)-H_i(t))-(H_j(t-\tau )-H_i(t-\tau))
\end{equation}
where $\theta_\Delta$ represents the changes in clock offset between $v_i$ and $v_j$ over time $\tau$, i.e., the clock offset increment. Thus, $\varphi_i^j$ can be calculated by any two clock offset values and the corresponding $\tau$. The relative clock skew in (\ref{equ:3}) is a direct calculation.

As discussed in the preceding literature review, previous studies have focused on joint clock skew-offset estimation or have employed clock offset estimates to generate clock skew estimates. Thus, the clock offset and clock skew estimates are closely related in current time synchronization algorithms. However, the clock skew can be estimated from (\ref{equ:3}) based only on the value of $\theta_\Delta$.

\subsection{The Proposed Multiple One-way Broadcast Model}
The multiple one-way broadcast model is designed for the clock skew MLE proposed later in this study. Upon initiation, a synchronization process is generated at an arbitrary node $v_i$, which broadcasts a group of packets ($N$ in number) over a time interval (synchronization period) $T_b$. Then $N$ pairs of timestamps are produced between $v_i$ and the receiver. Based on the timestamps, the proposed clock skew MLE is derived and implemented.

\begin{figure}[!htb]
\centering
\includegraphics[scale=1.2]{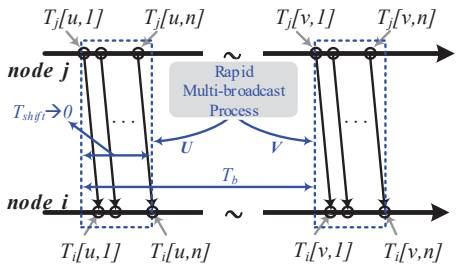}
\caption{The multiple one-way broadcast model showing the timestamps $T_i$ and $T_j$ created in $U$ and $V$.}
\label{FIG1_19-TIE-1783}
\end{figure}

The multiple one-way broadcast model is illustrated in Fig. \ref{FIG1_19-TIE-1783}. It includes two multi-broadcast processes: $U$ and $V$. Variable $T_{shift}$ is the time interval from the first packet to the last packet in the same multi-broadcast process. That is, $T_{shift}$ is the time cost description of the multi-broadcast process. The MAC-layer timestamps, $\{T_i [u,n]\}_{n=1}^N $ and $\{T_i [v,n]\}_{n=1}^N$, are generated at node $v_i$ while the MAC-layer time stamps, $\{T_j [u,n]\}_{n=1}^N$ and $\{T_j [v,n]\}_{n=1}^N$, are generated at $v_j$. It is clear that the multiple one-way broadcast model does not require topology management, and the sender does not need to maintain the neighbor's information while the receiver only needs to handle the neighbor's time information.

\vspace{0.3cm}

Considering the changes in the clock offset due to clock drift, the multi-broadcast process should be completed as soon as possible to obtain the fixed clock offset during the same multi-broadcast process. That is, it is required that the clock offset be a constant in the time interval $T_{shift}$. Therefore, $T_{shift}$ should be as small as possible, i.e., $T_{shift}\rightarrow0$, and is required to satisfy the inequality
\begin{equation}\label{equ:20}
  T_{shift}< \frac{1}{\varphi_i^j\times f_s}
\end{equation}
where $f_s$ is the nominal frequency of $H(t)$. The inequality indicates that the relative clock offset increment between $v_i$ and $v_j$ is smaller than the clock granularity of $H(t)$, i.e., $\theta_\Delta(\tau)<1/f_s$ where $\tau=T_{shift}$.

\subsection{Actual Experimental Delay}
The delays in one-way broadcasting strongly depend on hardware and software implementation \cite{24}. To test the actual experimental delay, MAC-layer timestamps based on the start-of-frame delimiter (SFD) interrupt (Section IV) is designed to measure the delay. Specifically, a single interrupt task experiment and a multi-interrupt experiment are designed.

The experimental results indicate that delay $D$ in one-way broadcast is composed of two portions
\begin{equation}\label{equ:9}
  D=D_{unc}+D_{var}=D_{unc}+(D_{fixed}+d)
\end{equation}
where $D_{var}$ is defined as the variable delay and $D_{unc}$ is defined as the uncertain delay; and $D_{fixed}$ and $d$ are the fixed portion and the variable portion of $D_{var}$, respectively \cite{27,34,40}. The mean of $D_{var}$ can be calculated as an estimation of $D_{fixed}$, i.e., $D_{fixed}=\bar{D}_{var}$.

The delay $D_{unc}$ may be caused by interrupt response blocking, which is inevitable in multi-interrupt system due to multilevel interrupt priority levels and interrupt nesting. The SFD interrupt response will be hung up by the processor when a higher priority interrupt request is generated, and it will be blocked when the higher priority request is responding.

\emph{\textbf{1) The variable delay}}

\begin{figure}[!htb]
\centering
\includegraphics[scale=0.6]{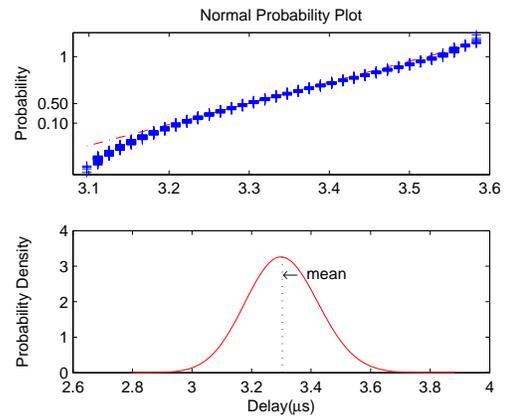}
\vspace{-0.2cm}
\caption{Normal probability plot and probability density of the variable delay $D_{var}$. If $D_{var}$ has a normal distribution, then they should tend to fall along the reference line in the normal probability plot.}
\label{FIG2_19-TIE-1783}
\end{figure}

A single interrupt (SFD interrupt) task testbed was designed to test $D_{var}$ and more than 500,000 data points were collected. The probability density is presented in the lower panel of Fig. \ref{FIG2_19-TIE-1783}. The maximum delay is approximately 3.6 {\textmu s}; the mean and the standard deviation of delay is about $3.3$ {\textmu s} and $0.072$, respectively. The confidence interval (95\%) of the mean of $D_var$ is 3.3026-3.3033 {\textmu s} and 0.07-0.0704 for the standard deviation.

Using a t-test to test the distribution of $D_{var}$, it is a normal distribution at 99.99\% confidence level. The normal probability plot of $D_{var}$ is presented in the upper panel of Fig. \ref{FIG2_19-TIE-1783}. The plot indicates that $D_{var}$ (and $d$) follows a normal distribution. However, both the mean and standard deviation are unknown for the MLE derivation.

\vspace{0.3cm}

\emph{\textbf{2) The uncertain delay}}

The multi-interrupt tasks experiment is designed to test $D_{unc}$. Three interrupt sources were set for the testbed, and three different priority levels were configured for the test sequence, i.e., equal priority level for all the interrupt sources, and lowest or highest priority levels for the SFD interrupt.

\vspace{0.8cm}
\begin{table}[htbp]
 \centering
 \captionsetup{justification=centering}
 \caption{\\Summary of the experimental results. Respectively, the 95\% confidence interval of the mean and standard deviation of $D_{var}$ are 3.3167-3.3176 {\textmu s} and 0.0668- 0.0674 in Equal, 3.3247-3.3254 {\textmu s} and 0.0665-0.0671 in Lowest, 3.3109-3.3117 {\textmu s} and 0.0701-0.0707 in Highest, respectively.}{\label{tab:2}}
 \begin{tabular}{ccccccc}
  \toprule
  \toprule
        SFD               &  \multicolumn{3}{c}{ \textbf{Variable delay}}&  \multicolumn{2}{c}{ \textbf{Uncertain delay} }  \\
     \tabincell{c}{Interrupt\\priority}
                          &\tabincell{c}{Probability}
                                               &\tabincell{c}{Mean\\({\textmu s})}
                                                                &STD            &\tabincell{c}{Probability}
                                                                &\tabincell{c}{Maximal\\({\textmu s})} \\
  \midrule
        Equal	          &0.8632	           &3.317	 	     &0.0671            &0.1368                &909	     \\
        Lowest	          &0.9232	           &3.325	   	     &0.0667	        &0.0768	                &732	     \\
        Highest	          &0.9933	           &3.311	   	     &0.0704	        &0.0067	                &909	      \\
  \bottomrule
  \bottomrule
 \end{tabular}
\end{table}

\begin{figure}[!htb]
\centering
\includegraphics[scale=0.6]{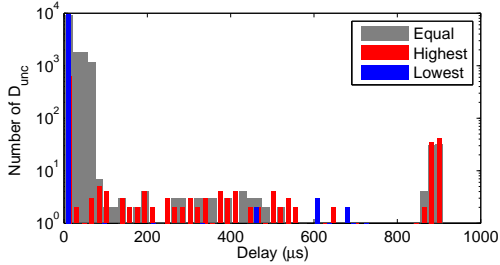}
\vspace{-0.2cm}
\caption{Actual distribution of $D_{unc}$ under different SFD interrupt priority levels setting. Delay $D_{unc}$ is larger than $D_{var}$.}
\label{FIG3_19-TIE-1783}
\end{figure}

Delay $D_{unc}$ looks like impulse noise or is an outlier to $D_{var}$, which is a sudden disturbance that occurs randomly in time and is an almost instantaneous (impulse-like) sharp delay.
Based on the variable delays shown in Fig. \ref{FIG2_19-TIE-1783}, we use the PauTa Criterion ($3\sigma$) to detect $D_{unc}$ from the observations. The summary of the experimental results (more than 300,000 data points were collected) is presented in Table \ref{tab:2}. The probability columns indicate the probability of $D_{var}$ and $D_{unc}$, respectively. This is about 0.1368 (maximal probability) for $D_{unc}$ in an equal priority level setting test, and about 0.9933 (maximal probability) for $D_{var}$ in a highest priority level setting test. The actual distributions of $D_{unc}$ are presented in Fig. \ref{FIG3_19-TIE-1783}.

The experimental results demonstrate that $D_{unc}$ is unavoidable, even with its very low probability, which may result into the clock skew estimation being incorrect and the clock skew cannot be ignored.

\vspace{-0.2cm}
\vspace{0.3cm}

\section{The MLE for Clock Skew}
Many distribution functions are used to model $d$ and the exponential model \cite{27,32,40,41,42} and Gaussian \cite{22,27,43,44} are popular used in existing clock skew MLE. Based on (\ref{equ:9}) and the experimental results in Fig. \ref{FIG2_19-TIE-1783}, $D_{var}$ is Gaussian distributed and it is possible to detect $D_{unc}$ from $D$ using the PauTa Criterion. The clock skew MLE is derived based on the Gaussian model, and only $D{var}$ is considered in this section.

\subsection{The Proposed Clock Skew Estimation}
Parameters $\{u[n]\}_{n=1}^N$ are the observation set for $U$, and $\{v[n]\}_{n=1}^N $ are the observation set for $V$. Specifically, $\{u[n],v[n]\}_{n=1}^N$ are given by
\begin{equation}\label{equ:5}
  u[n]\triangleq T_i[u,n]-T_j[u,n]=D_{fixed}+\theta_u+d[u,n],
\end{equation}
\begin{equation}\label{equ:6}
  v[n]\triangleq T_i[v,n]-T_j[v,n]=D_{fixed}+\theta_u+\theta_\Delta+d[v,n]
\end{equation}
where $ d:\{d[v,n],d[u,n]\}_{n=1}^N $ are denoted as the variable components of $D_{var}$ in $U$ and $V$.

Variable $\theta_u$ are denoted as the clock offset at $U$, and $\theta_\Delta$ are denoted as the clock offset increment from $U$ to $V$. We define $\hat{\theta}_\Delta=v[n]-u[n]$, and $\hat{\tau}=T_j[v,n]-T_j[u,n]$, where $n\in(0,N)$. Based on (\ref{equ:3}), if $\hat{\varphi}_i^j=\hat{\theta}_\Delta/ \hat{\tau}$ is the clock skew estimate, then $\hat{\varphi}_i^j$ is
\begin{equation}\label{equ:7}
  \hat{\varphi}_i^j[n]=(v[n]-u[n])/(T_j[v,n]-T_j[u,n])
\end{equation}
where $\hat{\varphi}_i^j$ is independent of the clock offset $\theta_u$, and the estimation errors depend completely on $\hat{\theta}_\Delta$. Based on (\ref{equ:5}) and (\ref{equ:6}), $\hat{\theta}_\Delta$ is rewritten as
\begin{equation}\label{equ:8}
  \hat{\theta}_\Delta=v[n]-u[n]=\theta_\Delta+d[v,n]-d[u,n].
\end{equation}

\subsection{MLE of Clock Skew}
The Gaussian distribution $N(\alpha,\sigma^2)$ is used to model variables $d:\{d[v,n],d[u,n]\}_{n=1}^N$, where $\alpha$ is 0, and $\sigma$ is unknown. To generate the MLE for $\hat{\varphi}_i^j$, we instead determine the MLE of $\hat{\theta}_\Delta$ based on (\ref{equ:7}) and (\ref{equ:8}). Variables $Z:\{z[n]\}_{n=1}^N$ are defined as $z[n]=d[v,n]-d[u,n], n\in(1,N)$. Note that $Z$ is a function of $d$. Then (\ref{equ:8}) can be rewritten as
\begin{equation}\label{equ:19}
  z[n]=v[n]-u[n]-\theta_\Delta.
\end{equation}

We define $P$ as
$$P:p[n]=v[n]-u[n], n\in(1,N)$$
where $p[n]$ are independent, the PDF of $Z$ can be derived using the convolution formula.

The PDF of random variables $d$ is $N(\alpha,\sigma^2)$, which is
\begin{equation}\label{equ:10}
  f(x)=\frac{1}{\sqrt{2\pi}\sigma} \exp(-\frac{(x-\alpha)^2}{2\sigma^2 }).
\end{equation}

The PDF of $Z$ is written as
\begin{equation}\label{equ:11}
  f(z)=\frac{1}{2\sqrt{\pi}\sigma} \exp(-\frac{z^2}{4\sigma^2 }).
\end{equation}

The likelihood function for $(\theta_\Delta, \alpha, \sigma^2)$, based (\ref{equ:9}) and (\ref{equ:11}), is
\begin{equation}\label{equ:12}
    \begin{aligned}
        L(P;\theta_\Delta,\alpha,& \sigma^2 )\\
                                 & =\frac{1}{(2\sqrt{\pi}\sigma)^N} \exp(-\frac{\sum_{n=1}^N(p[n]-\theta_\Delta)^2}{4\sigma^2 }).
    \end{aligned}
\end{equation}

Differentiating the log-likelihood function yields
\begin{equation}\label{equ:13}
    \frac {\partial \ln L(P;\theta_\Delta,\alpha,\sigma^2 )}{\partial \theta_\Delta}=\frac{1}{2\sigma^2} \sum_{n=1}^N(p[n]-\theta_\Delta).
\end{equation}

Therefore, the MLE of  $\hat{\theta}_\Delta$ is
\begin{equation}\label{equ:14}
     \begin{aligned}
        \hat{\theta}_{\Delta(MLE)} &=\arg{ \max{[\ln L(P;\theta_\Delta,\alpha,\sigma^2 )]} }\\
                                   &=\frac{\sum_{n=1}^N p[n]}{N}=\bar{P}.
    \end{aligned}
\end{equation}

Accordingly, the MLE of $\hat{\varphi}_i^j$ is
\begin{equation}\label{equ:15}
 \hat{\varphi}_{i(MLE)}^j=\frac{\bar{P}}{\hat{\tau}}
\end{equation}
where $\hat{\tau}$ is the estimate of the actual synchronization interval.

\subsection{CRLB}
The regularity condition of the CRLB holds because (\ref{equ:13}) is finite, and its expected value is zero. Thus, the CRLB for $\hat{\theta}_{\Delta(MLE)}$ can be obtained by differentiating (\ref{equ:13}) as
\begin{equation}\label{equ:16}
 \frac{\partial^2  \ln L(P;\theta_\Delta,\alpha,\sigma^2 )}{\partial \theta_\Delta^2 }=-\frac{N}{2\sigma^2 }.
\end{equation}

Therefore, the CRLB of $\hat{\theta}_{\Delta(MLE)}$ is
\begin{equation}\label{equ:17}
  \begin{aligned}
    \rm{var}(\hat{\theta}_{\Delta(MLE)} )&\geq -E{[\frac{\partial^2 \ln L(P;\theta_\Delta,\alpha,\sigma^2 )}{\partial \theta_\Delta^2}]}^{-1}\\
                                    &=\frac{2\sigma^2}{N}.
  \end{aligned}
\end{equation}

The variance of $\hat{\theta}_{\Delta(MLE)}$ approaches zero as $N$ increases, and is proportional to $\sigma^2$. Based on (\ref{equ:15}), its CRLB is
\begin{equation}\label{equ:18}
     {\rm{var}}(\hat{\varphi}_{i(MLE)}^j )\geq \frac{2\sigma^2}{\hat{\tau} N}.
\end{equation}

The MLE for clock skew estimation can be calculated using (\ref{equ:15}). The value of $\hat{\varphi}_{i(MLE)}^j$ strongly depends on the value of $\hat{\tau}$. Parameter $\hat{\theta}_{\Delta(e)}$ is the estimation error of $\hat{\theta}_{\Delta(MLE)}$, which is consistent whether using $\hat{\tau}_1$ or $\hat{\tau}_2$. Considering different broadcast intervals, e.g., ${\tau}_1$ and ${\tau}_2$ (${\tau}_1>{\tau}_2$), because of that $\hat{\varphi}_{i(MLE)}^j$ is inversely proportional to $\tau$, thus ${(\hat{\theta}_{\Delta(e)})}/{\tau_1}<{(\hat{\theta}_{\Delta(e)}/{\tau_2})}$. According to (\ref{equ:18}), the larger $\tau$ will leads to a smaller value of ${\hat{\theta}_{\Delta(e)}}/{\tau}$ for $\hat{\varphi}_{i(MLE)}^j$. Thus $\tau$ should be restrained when the proposed MLE is employed by a completely distributed time synchronization algorithm. In addition, the convergence time, clock drifting, and clock skew estimation errors should be carefully balanced.

\section{Implementation}

As the CRLB in (\ref{equ:18}) demonstrates that the accuracy clock skew MLE is affected by the number of timestamps, the synchronization period, and the distribution of delay ($N$, $\tau$, and $\sigma$). Hence, reasonable implementation of the MLE should consider the following problems:

1) The estimation error caused by $D_{unc}$. As discussed in Section III, $D_{unc}$ is not Gaussian and is much larger than $D_{var}$. If $D_{unc}$ is generated to the timestamp, the proposed $\hat{\varphi}_{i(MLE)}^j$ will be an unrealistic estimation.

2) The larger synchronization period $\tau$ may lead to better clock skew estimation in (\ref{equ:15}). However, the shorter the resynchronization (clock offset compensation) interval, the more accurate the synchronization. Therefore, it cannot simply extend the synchronization interval to meet the CRLB of MLE.

\begin{figure}[!htb]
\centering
\includegraphics[scale=0.8]{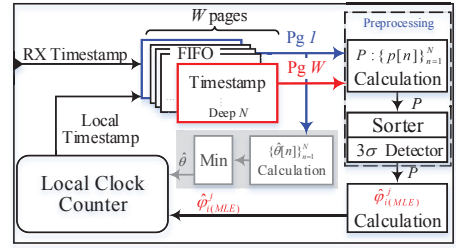}
\vspace{-0.2cm}
\caption{Implementation of the proposed MLE and time synchronization algorithm. Parameter $W\geq2$ is the length of the MLE sliding window.}
\label{FIG4_19-TIE-1783}
\end{figure}
\vspace{0.5cm}

As shown in Fig. \ref{FIG4_19-TIE-1783}, a reliable implementation for the proposed MLE is designed to address the problems outlined earlier. It employs three portions to implement the clock skew MLE (FIFO), preprocessing model, and $\hat{\varphi}_{i(MLE)}^j$ calculation.

The FIFO is designed to create a large $\tau$ of $\hat{\varphi}_{i(MLE)}^j$. It has $W\geq2$ page memories and a depth (multiple numbers) $N$ in each page. The memories are used to buffer the timestamp observations (e.g., ${T_i[u,n],T_j[u,n]}_{n=1}^N$) in sequence. A timestamp of $W$ multiple one-way broadcasting processes can be buffered by the FIFO, i.e., $N\times W$ timestamps. In addition, a sliding window is generated for the clock skew MLE in which the maximal length of the sliding window is $W$. Specifically, it uses the latest timestamp observations (Pg $1$) and the $W$-th from last (i.e.,Pg $W$) to calculate the $\hat{\varphi}_{i(MLE)}^j$. Parameter $\tau$ in (\ref{equ:15}) is $(W-1)$ times the synchronization period $T_b$. To ensure fast convergence of the MLE, the length of sliding window is initialized to 2 and then incremented until it equals $W$.

The uncertain delay $D_{unc}$ can be removed from the observations $P$ for $\hat{\varphi}_{i(MLE)}^j$ calculation in the preprocessing phase. First, $P$ is sorted in ascending order. Then the 3$\sigma$ detector is used to filter $D_{unc}$ in $P$. Considering the very low probability of $D_{unc}$, the possible $D_{unc}$ samples are moved to the end of $P$, and then $\hat{\sigma}$ can be calculated from the front observations, e.g., $\hat{\sigma}[k]=$std$\{ p[n]_{n=1}^{k-1}\}, N/2<k\leq N$.

\section{Evaluation and Discussions}
In this section, we evaluate the MLE in our \emph{synchronous sensing wireless sensor (SSWS)} testbed. {The experimental results for the proposed MLE are compared to the LR in \cite{22,24,25}, the direct estimator in \cite{28,29,30,31}, and the KF clock skew estimation in \cite{KF_2}.} All experiments in this study were conducted indoors at a slowly changing temperature, with a temperature range of roughly 0-3$^\circ C$.

\subsection{Testbed Setup and Error Measurement}
\begin{figure}[!htb]
\centering
\includegraphics[scale=0.8]{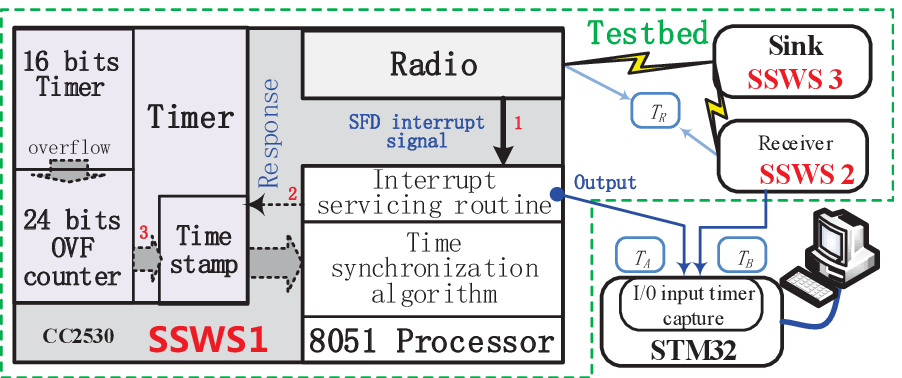}
\vspace{-0.1cm}
\caption{{The \emph{SSWS} and error measurement testbed. The SFD interrupt signal is used to create timestamps, and output for error measurement.}}
\label{FIG5_19-TIE-1783}
\end{figure}

The prototype system was built around 26 \emph{SSWSs} and designed based on the system-on-chip (SoC) wireless transmission chip CC2530. The rectangle in the left of Fig. \ref{FIG5_19-TIE-1783} indicates the structure of the \emph{SSWS}, and the portion in the blue dotted box indicates the structure of the testbed. The external 32 MHz crystal oscillator is set as the system clock source, and its frequency offset may be up to 50000 parts per billion (PPB), i.e., 50 {\textmu s} of clock offset increment per second. The MAC Timer 2 (which has a 16 bits timer and a 24 bits overflow counter) of the CC2530 is employed to generate the time notion for the time synchronization algorithm. Timer 2 works in up mode, and the clock granularity is determined by the overflow period of the up-counter of the timer.

\vspace{0.5cm}

The SFD interrupt handling is used to generate MAC-Layer timestamps. The SFD interrupt request is triggered once the SFD byte has been transmitted or received by radio. Then the timestamp is created during the interrupt service phase. The main delay of the timestamp is the interrupt handling time and the timestamp response time, as delay 1, 2 and 3 illustrate in Fig. \ref{FIG5_19-TIE-1783}. The different values of the timestamp delays between the sender and receiver present typical statistical characteristics and depend on the processing rate of the processor and the interrupt priority level.

To calculate the synchronization error, all the nodes should generate the timestamp at a same moment. Similar to \cite{25}, a sink-node, which is connected to a PC via a serial port, is employed to transmit the testing packet periodically at 10 second intervals. Once the test node receives a packet, it creates a timestamp immediately. All the test nodes upload their timestamp packets to the PC via the sink-node. The synchronization error is calculated on the PC using pairs of timestamps.

A testbed is built to test the measurement error. As shown in Fig. \ref{FIG5_19-TIE-1783}, if the packet arrives at different receivers at the same moment $T_R$, then the received SFD interrupt request signal will be output immediately. If the output signals are captured by STM32 at moment $T_A$ and $T_B$, then the measuring error is estimated as $E_{mea}=|T_A-T_B|$.

\begin{figure}[!htb]
\centering
\includegraphics[scale=0.6]{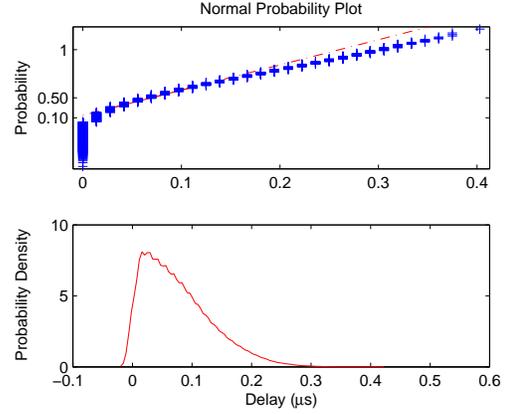}
\vspace{-0.2cm}
\caption{The normal probability plot and the probability density of the measuring error estimations $E_{mea}$. The $E_{mea}$ is normally distributed and at its 95\% confidence interval of the mean is 0.0756-0.0763 {\textmu s} and the variance is 0.0572-0.0578.}
\label{FIG6_19-TIE-1783}
\end{figure}

To measure $E_{mea}$, a timer of STM32 is set as an input capture model and its clock source is set as 72 MHz. The timer captures the rising edges of the output signals and records the corresponding count values. More than 500,000 data points were collected, and the probability density of $E_{mea}$ is presented in the lower panel of Fig. \ref{FIG6_19-TIE-1783}. The mean of $E_{mea}$ is about 0.0759 {\textmu s}; the maximum value is about 0.4028 {\textmu s}; the variance is about 0.0033. We used the t-test to test $E_{mea}$, and it shows that the $E_{mea}$ is a normal distribution with 99.99\% confidence. The normal probability plot is shown in the upper panel of Fig. \ref{FIG6_19-TIE-1783}.

Hence, once the clock granularity of time notion is greater than the maximum value of $E_{mea}$, then the $E_{mea}$ have no effect on the measuring precision, and the measuring error is in the tolerable range.

\vspace{0.5cm}

\subsection{Clock Skew Estimate Experimental Results}
A star network based on 26 \emph{SSWS} nodes was built to evaluate the performance of the proposed MLE in which the central node is set as a reference, and another 25 nodes are set as receivers. The reference broadcasts the time information periodically, and the receivers estimate the relative clock skew when they receive the time information. We report the statistical characteristics of the experimental results in this section.

\emph{\textbf{LR}}: The periodic one-way broadcast model is used. The period $T_b$ is 30 seconds (120 packets per hour) and the regression table size is 8. The LR converges after the 8th broadcast (approximately 240 seconds).

\emph{\textbf{MLE}}: The periodic multiple one-way broadcast model is used and the period $T_b$ is 200 seconds (90 packets per hour), and the multiple number $N$ is 5, and sliding window length $W$ is 2 (i.e., $\tau=T_b$ ). The MLE converges after the second round of broadcast (approximate 230 seconds).

\emph{\textbf{Direct estimator}}: The implementation is the same as LR. It costs up to hundreds of synchronization periods to complete consensus in a multiple hop network \cite{28,29}, and its performance degrades over longer synchronization periods \cite{28}.

\emph{\textbf{KF}}: In \cite{KF_2}, a lot of packets are required to be transmitted in a short time (e.g., 16-64 message per second and continuous for 4 minutes). For the KF, this will result in rapid energy consumption and a high probability communication collision. Unfortunately, the implementation of LR cannot meet the KF due to the changes in clock drift. The optimized implementation of KF is designed based on the multiple one-way broadcast model, where $N$ is 5 and the period is 30 seconds (600 packets per hour). The sliding window length is 8 for KF. The KF converges after the 7-th round, i.e., 210 seconds.

We performed an over 16 hours run for LR, more than 9 hours run for direct estimator, and more than 13 hours run for MLE and KF. The statistical average experimental results for absolute clock skew estimation error are reported in the following figures.

\begin{figure}[!htb]
\centering
\includegraphics[scale=0.6]{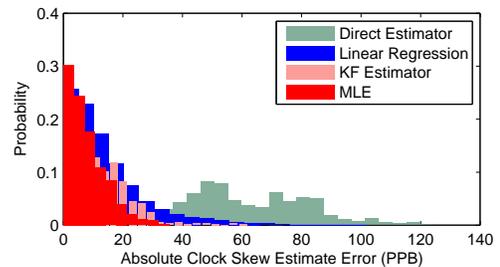}
\caption{The probability of absolute clock skew estimation errors. The maximum error is about 139 PPB in direct estimator, 102 PPB in LR, 62 PPB in KF, and 36 PPB in MLE.}
\label{FIG7_19-TIE-1783}
\end{figure}
\vspace{0.3cm}

\begin{figure}[!htb]
\centering
\includegraphics[scale=0.6]{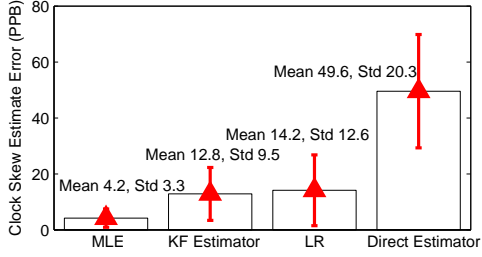}
\vspace{-0.2cm}
\caption{The absolute clock skew estimation error bar. The bar indicates the standard deviation over time.}
\label{FIG8_19-TIE-1783}
\end{figure}
\vspace{0.3cm}

Delay ($D_{var}$ and $D_{unc}$) is the main estimation error of clock skew estimation, and it goes directly to the direct estimator but is effectively minimized in MLE, LR and KF. The actual distributions of clock skew estimation errors are demonstrated in Fig. \ref{FIG7_19-TIE-1783}. The distribution is flatter and the estimation errors deviate from zero in the direct estimator, while other distributions (probabilities) are steeper and closer to zero. It can be seen that MLE outperforms the other algorithms, and the value and probability of maximal clock skew estimation error in MLE are the least.

The error bar of the absolute clock skew estimation is plotted in Fig. \ref{FIG8_19-TIE-1783}. The error bars also show the confidence intervals of the clock skew estimation errors, and the proposed MLE contributes the lowest estimate error for the clock skew estimation. The results also show that MLE is robust and the mean and standard deviation are smaller than that of other methods.

It can be seen from the experimental results that our method achieves better convergence rate and clock skew estimation than all the other methods in comparison. The average clock skew estimation errors of LR, KF and direct estimator are about 3-4 times, 2-3 times and 12-13 times to MLE, respectively. For the same broadcast period, e.g., 30 seconds, the convergence time is about 60 seconds in MLE and direct estimator, while 240 seconds in LR.

\subsection{Clock Skew Estimation Error vs. Synchronization Period}
For the time synchronization algorithm, the shorter the synchronization period is, the better synchronization the accuracy but the worse the energy efficiency becomes. There are different typical configurations to the synchronization interval in different standards, e.g., 30 second in ISA100.11a \cite{ISA-100.11a}, an 50 second in WirelessHART \cite{WirelessHART}. Equations (\ref{equ:15}) and (\ref{equ:18}) indicate that the low bound of clock skew estimation error is the proportion of synchronization period.. Here we evaluate the advantages of the proposed MLE in diverse synch period based on the testbed above. The statistical characteristics of the experimental results for all the nodes are reported.

\begin{figure}[!htb]
\centering
\includegraphics[scale=0.6]{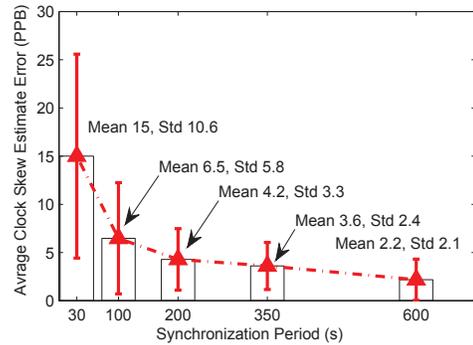}
\vspace{-0.2cm}
\caption{The absolute clock estimation error bar under diverse synch period. The clock skew estimation is more accurate under the larger synchronization interval; meanwhile, the total broadcast is less.}
\label{FIG9_19-TIE-1783}
\end{figure}

Figure \ref{FIG9_19-TIE-1783} shows the error bars of the MLE clock skew estimation under different synch periods. We report the experimental results of over 4 hours run for each synch period. It shows the confidence intervals of the clock skew estimation errors, and both mean and variance of the synchronization error are decreased as the synch period increasing. As the results show, the clock skew estimation improvement is fastest before the 350 second period. After that period, the decreasing tendency becomes slower. It also shows that the longer synch period contributes to higher precision for MLE theoretically.

\begin{figure}[!htb]
\centering
\includegraphics[scale=0.6]{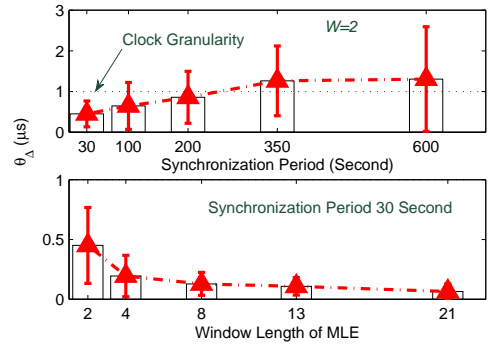}
\vspace{-0.1cm}
\caption{The maximum value of $\theta_\Delta$. The upper panel of the figure is the result of the MLE under diverse synchronization period, and when the sliding window length $W$ is 2. The lower panel of the figure is the result of the MLE in the proposed implementation, and $T_b$ is 30 seconds.}
\label{FIG10_19-TIE-1783}
\end{figure}

Considering changes in the environment and clock drift precisely, it must be considered that the clock skew is always changing, and a larger period may introduces more estimation errors \cite{28,38}. The error bars of $\theta_\Delta$ under the diverse sync periods are presented in the upper panel of Fig. \ref{FIG10_19-TIE-1783}. The time synchronization errors are higher for the longer synch periods \cite{28}. The period of a time synchronization algorithm that employs MLE cannot be too large, thus choosing an appropriate synchronization period is very important for the wireless network applications.

Therefore, the optimized implementation is very important to the MLE. As shown in Section IV, the FIFO provides an estimation sliding to the MLE, where $\tau$ is an integer multiple of $T_b$. Considering the 30 second synchronization period, if $W$ is 8, then $\tau$ for $\hat{\varphi}_{i(MLE)}^j$ is $30\times 7$ seconds. The error bars of $\theta_\Delta$ under diverse sliding window are shown in lower panel of Fig. \ref{FIG10_19-TIE-1783}. The time synchronization errors are decreased for the larger window length. The results indicate that the implementation in Fig. \ref{FIG4_19-TIE-1783} is better than the implementation with $W=2$ and $\tau=T_b$.

\vspace{0.5cm}

\vspace{0.5cm}

\subsection{Clock Skew Estimation Error vs. Uncertain Delay}
A star network with 4 \emph{SSWS} nodes is built, in which one node is set as a reference to broadcast packets, one is used to collect data, and two nodes runs MLE and LR separately. In the experiment, the uncertain delay $D_{unc}$ is about 200 {\textmu s}.

\begin{figure}[!htb]
\centering
\includegraphics[scale=0.6]{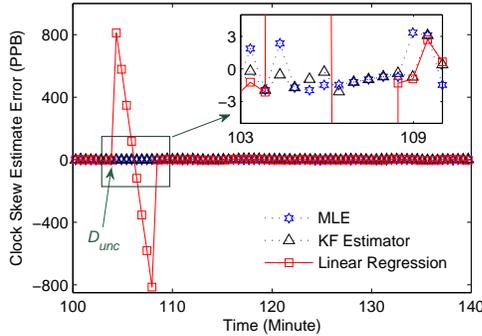}
\vspace{-0.2cm}
\caption{The clock skew estimation error at uncertain delay.}
\label{FIG11_19-TIE-1783}
\end{figure}

The uncertain delay may reduce the synchronization accuracy and may even cause the synchronization to fail. The experimental results in Fig. \ref{FIG11_19-TIE-1783} indicate that the uncertain delay will result in incorrect clock skew estimation to LR, and it costs about 8 synchronization periods (depends on the size of regression table) to recover the accurate estimation. While the clock skew estimation is not affected by $D_{unc}$ in MLE and KF, the experimental results show that $D_{unc}$ is removed from the estimation by the preprocessing model (in Fig. \ref{FIG4_19-TIE-1783}). The KF uses real-time variance to compensate for clock offset estimates, thus the effect of $D_{unc}$ is also not significant.

\vspace{-0.3cm}
\vspace{0.5cm}

\subsection{Clock Skew Estimation Error vs. Clock Granularity}

The clock granularity is different from the different applications, e.g., about 1 {\textmu s} to the industrial automation, about 30.5 {\textmu s} to the low-power sleeping. In this part, we evaluate the MLE at different clock granularity setting and discuss the effect of clock resolution on the MLE estimation error.
We configure the overflow period of up-counter of Timer 2 to generate clock sources with different resolutions, i.e., 0.125 {\textmu s}, 0.5 {\textmu s}, 1 {\textmu s}, 5 {\textmu s}, 10 {\textmu s} and 32 {\textmu s}. The sliding window length $W$ is 8 and the multiple number $N$ is 20 in the MLE and KF, and synchronization period is 30 seconds.

\begin{figure}[!htb]
\centering
\includegraphics[scale=0.6]{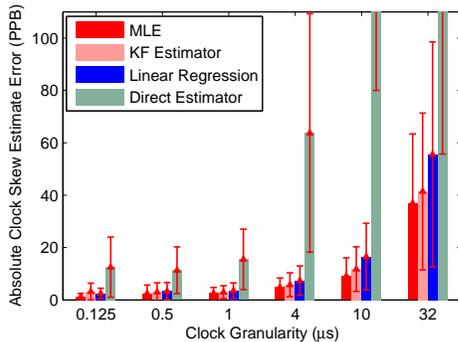}
\vspace{-0.2cm}
\caption{The absolute of clock skew estimation error at different clock granularity, i.e., 0.125 {\textmu s}, 0.5 {\textmu s}, 1 {\textmu s}, 5 {\textmu s}, 10 {\textmu s} and 32 {\textmu s}.}
\label{FIG12_19-TIE-1783}
\end{figure}

A master-slave testbed is built and more than 6 hours of experimental results are counted at different clock resolutions, and the statistical characteristics of clock skew estimation at slave node is shown in Fig. \ref{FIG12_19-TIE-1783}.

\vspace{-0.3cm}
\vspace{0.5cm}

The clock resolution constrains the lower bound of the time synchronization algorithm. It determines the minimum unit of the time counter and affects the actual distribution of delays in the timestamp. The fine-grained clock is more capable of picking up accurate delays. Taking the clock source of 32 {\textmu s} as example, if the actual delay is 3 {\textmu s}, then the time stamping error is about 3 {\textmu s} when the counter ignores the delay, or 29 {\textmu s} when counter picks up the delay.

Therefore, the coarse-grained clock will cause the larger estimation error. It can be seen from the experimental results that when the clock granularity becomes rough, the estimation errors of the algorithms are increasing. However, the estimation error in MLE is less and the increase rate is significantly slower than that of the comparison methods.

\subsection{MLE Clock Skew in Flooding Synchronization}
The implementation in Fig. \ref{FIG4_19-TIE-1783} also demonstrates a simple synchronization algorithm, in which a minimum function is used to create clock offset MLE (the shadow region): the timestamps in Pg $1$ are used to generate clock offset estimates $\{\hat{\theta}[n]\}_{n=1}^N$; then $\min\{\hat{\theta}[n]\}_{n=1}^N$ is used to compensate local clock, and he clock offset estimation error could be reduced to $D_{fixed}+\min(d)$. In this part, we use the rapid-flood that used in PulseSync to implement the MLE synchronization algorithm in multiple networks, and define the algorithm as MLE-PulseSync. In addition, a new flood time synchronization RMTS is proposed based on MLE in \cite{RMTS}. A 24 hops line network is built on 25 \emph{SSWS} nodes, i.e. \textbf{$\emph{R}\leftarrow1\leftarrow2\leftarrow\ldots\leftarrow24$}. Node \textbf{\emph{R}} is reference or root. The synchronization interval is 30 seconds for PulseSync, and is 50 seconds for MLE-PulseSync. We assumed that fixed delay $\hat{D}_{fixed}$ is 3 {\textmu s} and compensate it to the clock offset estimation. The local sync-error is defined as the time synchronization error between adjacent nodes, and the global sync-error is defined as the time synchronization error between any pair of nodes.

\begin{figure}[!htb]
\centering
\includegraphics[scale=0.6]{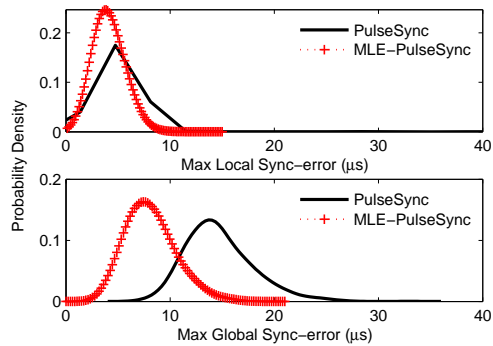}
\vspace{-0.2cm}
\caption{The probability density. The upper panel of the figure shows maximum local sync-error and the lower panel of the figure shows maximum global sync-error.}
\label{FIG13_19-TIE-1783}
\end{figure}

The upper panel in Fig. \ref{FIG13_19-TIE-1783} shows the maximum local sync-error probability density. The mean and standard deviation of maximal local sync-error (which are calculated when the clock skew estimation has been completely converged) are 4.09 {\textmu s} and 1.28 for MLE-PulseSync, and 5.04 {\textmu s} and 1.69 for PulseSync. The probability density of local sync-error in MLE-PulseSync is tighter than that of PulseSync.

The lower panel of Fig. \ref{FIG13_19-TIE-1783} shows the maximum global sync-error probability density. The mean and standard deviation of maximal global sync-error are 7.4 {\textmu s} and 2.32 for MLE-PulseSync, and 13.7 {\textmu s} and 3.22 for PulseSync. Obviously, the MLE-PulseSync contributes the better performances, it has lower synchronization errors and narrow error range. It means that the MLE-PulseSync may contribute to higher and more stable synchronization precision in large-scale WSN.

\section{Conclusion}
In this study, we propose a clock skew MLE based on the multiple one-way broadcast model and Gaussian distribution variable delay, and its CRLB is proportional to the synchronization period. In the implementation of the proposed MLE, a sliding window based on FIFO is designed to meet the accurate MLE at a reasonable synchronization period (e.g., a short resynchronization interval to create accurate synchronization), and a preprocessing (sorting and $3\sigma$ detecting) model is designed to remove possible uncertain delay.

The experimental results indicate that the proposed clock skew estimation has fast convergence, and is accurate for different synchronization periods and clock granularity.
Therefore, it is possible to compensate for clock skew and force all logical clocks, first, to run at almost the same speed, and then immediately establish accurate time synchronization with clock offset compensation. The proposed MLE provide an accurate time synchronization algorithm for WSNs. In addition, it may also be suitable for providing the time synchronization required in other wireless network applications, e.g., networked control, wireless monitoring and distributed measurement.

\ifCLASSOPTIONcaptionsoff
  \newpage
\fi

{
\scriptsize

\bibliographystyle{ieeetr}
\bibliography{TXT_19-TIE-1783}
}

\begin{IEEEbiography}[{\includegraphics[width=1in,height=1.25in,clip,keepaspectratio]{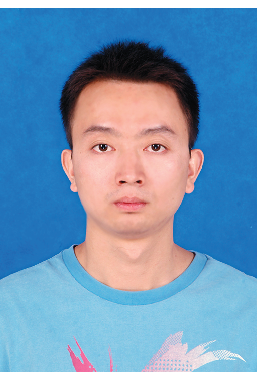}}]{Fanrong Shi}
 (S'18) received the B.E., M.E. and Ph.D. degrees in communication engineering, communication and information system, control science and engineering from Southwest University of Science and Technology (SWUST), Mianyang, China, in 2009, 2012, 2019, respectively. Since 2012, he is currently an Assistant Professor in information and communication engineering with the Department of School of Information Engineering at SWUST. His research interests include time synchronization and location in wireless sensor networks, wireless measurement, and signal acquisition.
\end{IEEEbiography}

\begin{IEEEbiography}[{\includegraphics[width=1in,height=1.25in,clip,keepaspectratio]{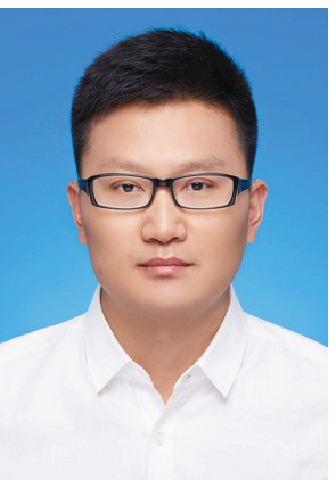}}]{Huailiang Li}
(M'18) received the B.E., M.E., and Ph.D. degrees in information engineering, information and communication engineering, earth exploration and information technology, from the Chengdu University of Technology, Chengdu, China, in 2007, 2010, and 2013, respectively.
Since 2013, he worked as Associate Professor or Professor with the Fundamental Science on Nuclear Wastes and Environmental Safety Laboratory, Southwest University of Science and Technology in Mianyang, Sichuan, China. He is currently a Professor in signal and information processing at the College of Geophysics, Chengdu University of Technology, Chengdu, China. His research interests include data acquisition, signal processing and wireless sensor network in geosciences.
\end{IEEEbiography}

\begin{IEEEbiography}[{\includegraphics[width=1in,height=1.25in,clip,keepaspectratio]{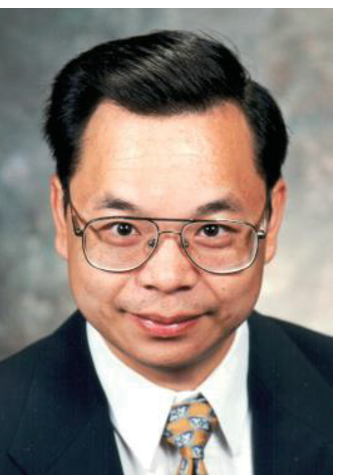}}]{Simon X. Yang}
(S'97, M'99, SM'08) received the B.Sc. degree in engineering physics from Beijing University, China in 1987, the first of two M.Sc.  degrees in biophysics from Chinese Academy of Sciences, Beijing, China in 1990, the second M.Sc. degree in electrical engineering from the University of Houston, USA in 1996, and the Ph.D. degree in electrical and computer engineering from the University of Alberta, Canada in 1999. Currently he is a Professor in systems and computer engineering and the Head of the Advanced Robotics and Intelligent Systems (ARIS) Laboratory at the University of Guelph in Canada. His research interests include intelligent systems, robotics, sensors and multi-sensor fusion, wireless sensor networks, control systems, and computational neuroscience. Prof. Yang serves as the Editor-in-Chief of International Journal of Robotics and Automation, and Associate Editor of IEEE Transactions on Cybernetics, and several other journals. He has involved in the organization of many conferences.
\end{IEEEbiography}

\begin{IEEEbiography}[{\includegraphics[width=1in,height=1.25in,clip,keepaspectratio]{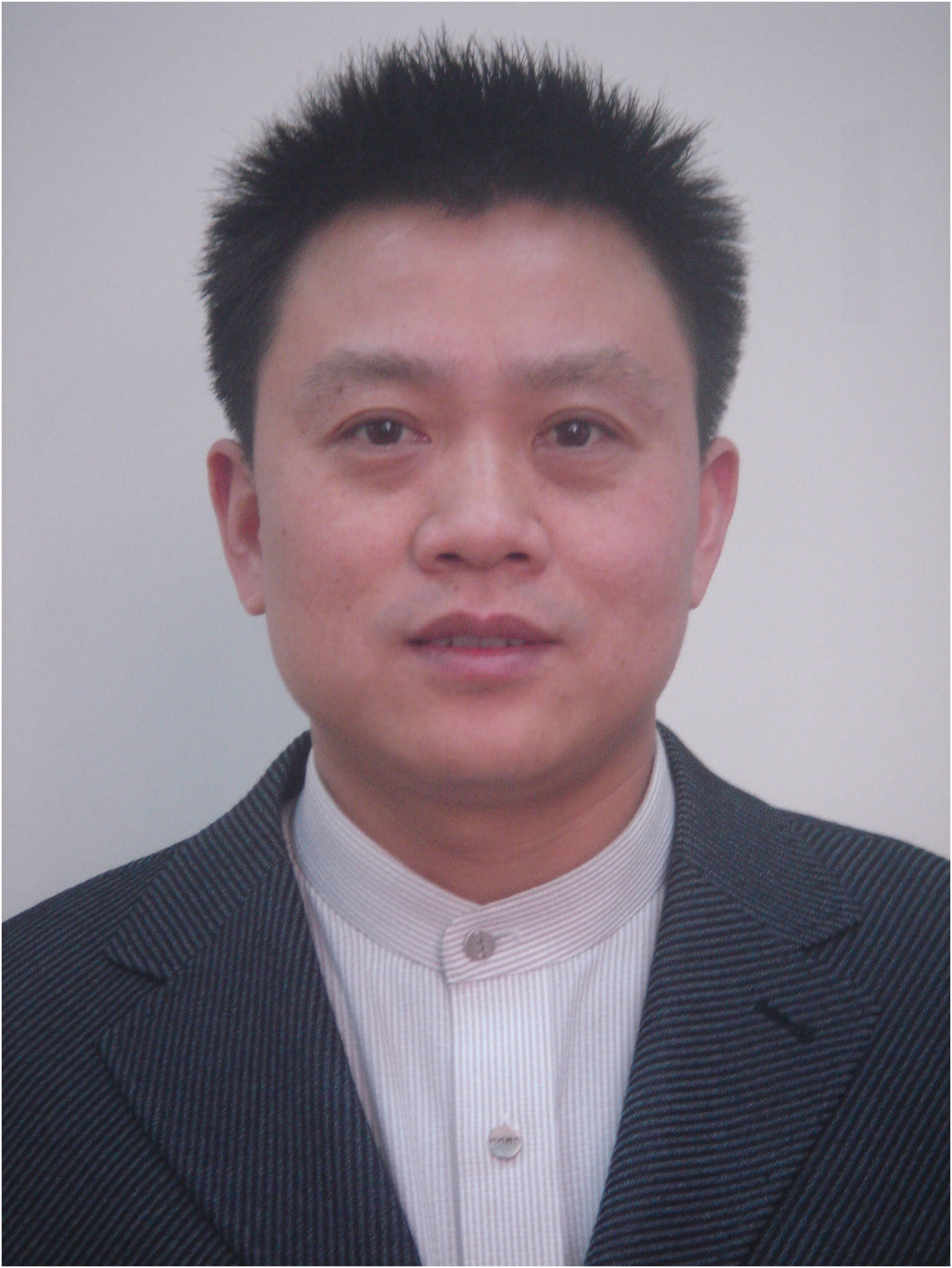}}]{Xianguo Tuo}
received the B.E., M.E. and Ph.D. degrees in radioactive geophysical exploration, radioactive geology and exploration, nuclear technology and applications from the Chengdu University of Technology in Chengdu, China, in 1988, 1993, 2001 respectively. From 2006 to 2007, he worked as visiting scholar at the School of Bioscience at University of Nottingham, UK. Since 2001, he is a Professor with College of Nuclear Technology and Automation Engineering, Chengdu University of Technology, China. Since 2012, he becomes a Professor with the School of National Defense Science and Technology, Southwest University of Science and Technology in Mianyang, Sichuan, China. Currently, he is a Professor in artificial intelligence and robot technology, and is the President of Sichuan University of  Science and Engineering, Zigong, Sichuan. Professor Tuo received The National Science Fund for Distinguished Young Scholars in 2011. Currently, his research interests are in detection of radiation, seismic exploration and specialized robots.
\end{IEEEbiography}

\begin{IEEEbiography}[{\includegraphics[width=1in,height=1.25in,clip,keepaspectratio]{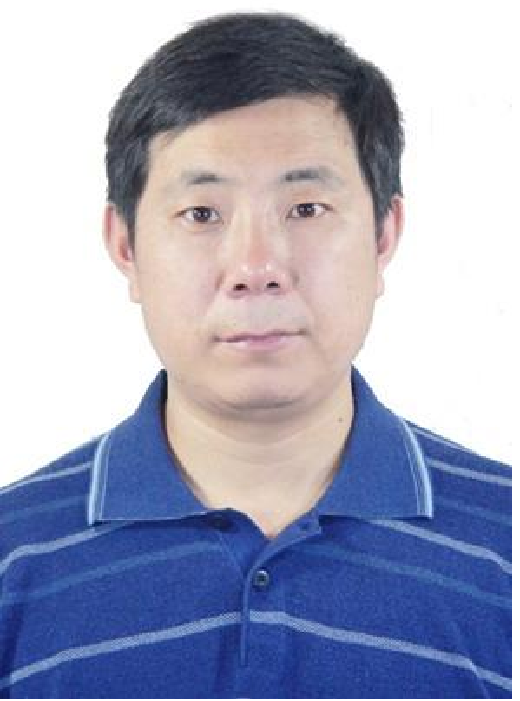}}]{Maosong Lin}
received the M.E. and Ph.D. degrees in Mechanical design and theory from Southwest Jiaotong University, Chengdu, China, in 1995, 2006. He is currently an Professor in Computer application with the Department of School of computer science and technology, Southwest University of Science and Technology. His research interests include image processing, computer vision.
\end{IEEEbiography}

\end{document}